\documentclass[graphicx]{article}

\usepackage{amsfonts}
\usepackage{amssymb}
\usepackage{longtable}
\usepackage{hyperref}
\usepackage{bm}
\usepackage[utf8]{inputenc}
\usepackage{graphicx}
\usepackage{array}
\usepackage{float}
\usepackage{sidecap}
\usepackage{authblk}
\DeclareGraphicsExtensions{.pdf,.eps,.jpg,.jpeg}

\newcommand{\bea}{\begin{eqnarray}}
\newcommand{\eea}{\end{eqnarray}}
\renewcommand{\vec}{\mathbf}

\title{Nonequilibrium transport equations and \textit{ab initio} study of
adsorption processes on carbon nanotubes}
\date{}
\author[1,2]{A.I.Vasylenko\thanks{vasylenko@icmp.lviv.ua}}
 
\affil[1]{Institute for Condensed Matter Physics National Academy of Science of Ukraine, 1, Sventsitskogo str., 79011 Lviv, Ukraine}
\affil[2]{NanoBioMedical Centre Adam Mickiewicz University, ul.Umultowska 85, PL 61614 Poznan, Poland}
\author[1]{M.V.Tokarchuk}

\author[2]{S.Jurga}

\begin{document}

\maketitle

\begin{abstract}
In a theoretical study of gas adsorption on carbon nanotubes (CNT) nonequilibrium processes of ionization, polarization, surface diffusion and desorption of atoms are considered self-consistently. The approach is based on Zubarev's method of nonequilibrium statistical operator and reaction-diffusion theory. The set of nonlinear transport equations are obtained for the chosen parameters of description: the average numbers of adsorbed atoms, ionized and polarized atoms in the electromagnetic field of CNT, and the average number of atoms desorbed from the CNT surface. \textit{Ab initio} simulations are conducted for a ``gas-single wall carbon nanotube'' system for gases of particular practical interest: He and NO. The obtained values of adsorption energy reveal preferable localization sites of absorbed He atoms as well as their dependency on adsorption distances. A significant effect of NO adsorption on CNT electronic properties is demonstrated. The effect of presence of vacancies on adsorption nature is analyzed. It is shown that under the influence of vacancy formation the CNT structure undergoes reconstruction that enables chemisorption of NO molecules.
\end{abstract}

\maketitle
\newpage
\section{Introduction}

The carbon nanotube (CNT) is considered a promising candidate for nanosensors since the time of its first application as gas detector \cite{1}.  Remarkable structural, electromagnetic and emission properties of CNT make them applicable as miniaturized sensors with a variety of sensing mechanisms: conductivity response, electrical discharge theory, chemical resistance, ionization sensors etc. \cite{2,3,4,5,ions}.
Many practical techniques have been developed for enhancement of CNT gas sensitivity \cite{COOH,ads1,ads2,Pt}. Despite of tremendous achievements in devising, understanding of some of the major underlying detection processes is still to be developed within a rigorous theoretical approach.
Complex analytical description must take into account the nonequilibrium processes of excitation, ionization and polarization of gas atoms in the field of CNT as well as the adsorption and desorption processes. Some of these processes are partially described within nonequilibrium theory of surface ionization \cite{ntsi1,ntsi2}. The mentioned theory enables one to obtain the rate of suface ionization of gases, which is the ratio of the current of desorbed atoms to the current of neutral atoms. The mentioned approach treats processes of adsorption as static ones and does not take into account the diffusion processes of adsorbed atoms on the surface.

Computational progress in description of gas adsorption on CNT is based mostly on the \textit{ab initio} technique \cite{5, ads1, ads2, Pt, abinit, abinit2}. Adsorption processes of several types of gas atoms and small molecules were investigated on pristine and doped SWCNT. It was found that Au-doped and Pt-doped SWCNT demonstrate enhanced sensitivity to SO2 and H2S, which relies on large charge transfer between CNT and adsorbed molecule resulting in a considerable change of electrical conductivity \cite{ads1,ads2,Pt}. Peng et al. \cite{abinit} and Wang et al. \cite{abinit2} found that B, N or Al-doped CNT gas sensors have enhanced sensitivity to CO and H2O in comparison to pristine CNT. It was also shown that response characteristics of sensors largely depend on the number of active sites, which strenghthen the response \cite{Pt, abinit2}.

Detection of some gases can be of particular practical interest. A recently developed Scanning Helium Ion Microscope \cite{HIM} proposes unrivaled imaging abilities, which in large part depend on the ability of detection of He. Nitric oxide is a bioproduct in almost all types of organisms, ranging from bacteria to  human cells \cite{NO1}. It is known as a cardiovascular signaling molecule \cite{NO}. Changes in NO levels can signify development of degenerative diseases such as Alzheimer’s that cause neuron death. It is also known that the exhaled NO level is markedly elevated in patients with asthma \cite{asthma}. Development of a highly sensitive device for NO detecting could have important applications in bioimaging and medical field.

The present work consists of two sections. In the first section we employ an analytical approach of the Zubarev nonequilibrium statistical operator and develop a set of generalized transport equations. These equations describe nonequilibrium transport of the chosen parameters of a ``gas -- electromagnetic field -- CNT'' system, which are the average numbers of neutral, ionized, adsorbed, diffusing on surface, and desorded atoms. In the second section we investigate adsorption processes by means of \textit{ab initio} computational technique and obtain adsorption energies and their dependencies on adsorption distances and atoms localizations. Based upon calculated band gaps we investigate the effect of adsorption of gases on electronic properties of CNT. The reconstruction of CNT surface, while vacancy is being formed is demonstrated as well as its drastic effect on nature of NO adsorption.

\section{Nonequilibrium transport equations of adsorption processes in a system ``gas--carbon nano\-tubes''}

Hamiltonian of a system of ``gas atoms-electromagnetic field-CNT'' can be presented as the following \cite{66}:
\bea\label{1.4.4}
&&\hat H = \int_0^\infty \hbar\omega\int\hat f^+(\vec R;\omega)\hat f(\vec R;\omega)d\omega d\vec R\\\nonumber
&&+\sum_{i=1}^{N}\frac{1}{2m_i}\left(\vec p_i+\frac{Z_ie}{c}\hat \vec a(\vec r_A+\vec r_i) \right)^2\\\nonumber
&&+\frac{1}{2}\int\hat\rho_A(\vec r)\hat\varphi_A(\vec r)d\vec r+\frac{1}{2}\int\hat\rho_A(\vec r)\hat\varphi(\vec r)d\vec r,
\eea
where $m_i, Z_ie, \vec r_i, \vec p_i$ stand for mass, charge, coordinate and momentum of the particles respectively, $c$ denotes speed of light.
\bea\label{1.4.5}
\hat{\varphi}_A(\vec r)=\int\frac{\hat\rho_A(\vec r')}{|\vec r-\vec r'|}d\vec r'
\eea
is a scalar potential of charged particles with  density
\bea\label{1.4.6}
\hat\rho_A(\vec r)=\sum_{i=1}^N Z_ie\delta(\vec r-\vec r_A-\vec r_i).
\eea
$\hat{ \varphi}(\vec r), \hat{\vec a}(\vec r)$ denote scalar and vector potentials of quantized electromagnetic field of CNT respectively, $\vec r = {r, \varphi, z}$. In Schr\"{o}dinger representation these potentials are related to corresponding components of electromagnetic fields as following:

\bea\label{1.4.7}
\hat{\vec a}(\vec r)=\int_0^\infty d\omega\frac{c}{i\omega}\hat{ E}^{\bot}(\vec r;\omega)+h.c.
\eea

\bea\label{1.4.8}
-\nabla\hat{\varphi}(\vec r)=\int_0^\infty d\omega\hat{E}^{\|}(\vec r;\omega)+h.c.
\eea

where

\bea\label{1.4.9}
\hat {E}^{\bot(\|)}(\vec r;\omega)=\int d\vec r'\delta^{\bot\|}(\vec r-\vec r')\hat{\vec E}(\vec r;\omega),
\eea

and
$$\delta^{\|}_{\alpha\beta}(\vec r)=-\nabla_{\alpha}\nabla_{\beta}\frac{1}{4\pi|\vec r|}, $$
$$
\delta^{\bot}_{\alpha\beta}(\vec r)=\delta_{\alpha\beta}\delta(\vec r)-\delta^{\|}_{\alpha\beta}(\vec r). $$

In Fourier representation the full electric field is related to the magnetic field and electric current by Maxwell equations
\bea\label{eqMax}
&&\nabla\times\hat{\vec E}(\vec r;\omega)=ik\hat{\vec H}(\vec r;\omega),\\\nonumber
&&\nabla\times\hat{\vec H}(\vec r;\omega)=-ik\hat{\vec E}(\vec r;\omega)+\frac{4\pi}{c}\hat{\vec J}(\vec r;\omega)
\eea
$$k=\omega/c$$
\bea\label{eqMax1}
\hat{\vec J}(\vec r;\omega)=\int d\vec R\delta(\vec r-\vec R)\hat{\vec J}(\vec r;\omega)=2\hat{\vec J}(R_n,\varphi,z;\omega)\delta(r-R_n)
\eea

According to (\ref{eqMax}), (\ref{eqMax1}) we obtain the relation
\bea\label{1.4.10}
\hat{\vec E}(\vec r;\omega)=i\frac{4\pi}{c}k\int d\vec R\stackrel{\leftrightarrow}{G}(\vec r,\vec R;\omega)\cdot\hat{\vec J}(\vec r;\omega),
\eea
where $\stackrel{\leftrightarrow}{G}(\vec r,\vec R;\omega)$ denotes Green's tensor of classical electromagnetic field of CNT. The components of Green's tensor satisfy the equations
\bea\label{1.4.11}
\sum_{\alpha=r,\varphi,z}(\nabla\times\nabla-k^2)_{z\alpha}G_{\alpha z}(\vec r,\vec R;\omega)=\delta(\vec r-\vec R).
\eea
Equations (\ref{1.4.10}) and (\ref{1.4.11}) enable us to obtain the important relation between the components of the electric field and the $z$-component of electric current
\bea\label{1.4.12}
\sum_{\alpha=r,\varphi,z}(\nabla\times\nabla-k^2)_{z\alpha}E_{\alpha z}(\vec r;\omega)=i\frac{4\pi}{c}k\hat J_z(\vec r;\omega).
\eea
Since the electric field can be decomposed by transverse and longitudinal terms
\bea\label{elF}
\hat{\vec E}(\vec r;\omega)=\hat{\vec E}^{\bot}(\vec r;\omega)+\hat{\vec E}^{\|}(\vec r;\omega),
\eea
which are related respectively to the vector and scalar potentials
\bea\label{vecP}
&&\hat{\vec E}^{\bot}(\vec r;\omega)=ik\hat{\vec a}(\vec r;\omega), \; \nabla\cdot\hat{\vec a}(\vec r;\omega)=0,\\\nonumber
&&\hat{\vec E}^{\|}(\vec r;\omega)=-\nabla\hat{\varphi}(\vec r;\omega),
\eea
equation for Green's tensor (\ref{1.4.11}) has a structure
\bea\label{1.4.13}
\sum_{\alpha=r,\varphi,z}(\nabla\times\nabla-k^2)_{z\alpha}\left[G^{\bot}_{\alpha z}(\vec r,\vec R;\omega)+G^{\|}_{\alpha z}(\vec r,\vec R;\omega)\right]=\delta(\vec r-\vec R).
\eea

In order to take into account polarization effects we decompose field operators $\hat{\varphi}(\vec r)$ and $\hat{\vec a}(\vec r)$ in Hamiltonian (\ref{1.4.4}) with respect to a center of mass of atoms $\vec r_A$,  keeping in mind that $\left[\vec p_i,\hat{\vec a}\right]=0$. Then we obtain for $H$
\bea\label{1.4.14}
\hat {H} = \hat {H}_f +\hat {H}_A+\hat {H}_{Af}^{\bot}+\hat {H}_{Af}^{\|},
\eea

\bea\label{1.4.15}
\hat {H}_f = \int_0^\infty \hbar\omega\int\hat {f}^+(\vec R;\omega)\hat {f}(\vec R;\omega)d\omega d\vec R,
\eea

\bea\label{1.4.16}
\hat {H}_A = \sum_{i=1}^N\frac{\vec p^2_i}{2m_i}+\sum_{i<j}\frac{Z_iZ_je^2}{|\vec r_i-\vec r_j|},
\eea

\bea\label{1.4.17}
\hat {H}^{\bot}_{Af}=-\sum_i \frac{Z_ie}{m_ic}\vec p_i\cdot\hat{\vec a}(\vec r_A)+\vec d\cdot\nabla\hat{\varphi}(\vec r_A),
\eea

\bea\label{1.4.18}
\hat {H}^{\|}_{Af}=\sum_i \frac{Z_i^2e^2}{m_ic^2}\hat{\vec a}^2(\vec r_A),
\eea

$\vec d = \sum_i Z_ie\vec r_i$ determines the operator of the electric dipole momentum of atom subsystem.

Therefore, gas atoms in quantum electromagnetic field of CNT \cite{66} after being polarized (effective dipole moment of atom is created) interact with the positively charged surface of CNT.  Ion-dipole attractive interaction appears between polarized atoms with effective dipole moment $d_{ef}$ and the charged surface of CNT

\bea\label{1.4.19}
V_{ia} = -\frac{Z_i^{ef}ed_{ef}}{r^2_{ia}}\cos(\theta).
\eea

As a consequence of this interaction a polarized atom may approach the surface at an electron tunneling distance to the nearest ion at the surface. A polarized atom therefore is adsorbed at the surface as an ion dipole. The ionized atoms adsorbed at the surface interact between themselves through effective dipole-dipole interaction, which creates diffusion processes. Theoretical description of adsorption and desorption processes should take into account all of the mentioned processes self-consistently. Hence, a wave function of nonequilibrium state of atoms should contain wave functions of the ground state  $\Psi_0(\vec r;t)$, the polarized state $\Psi_p(\vec r;t)$ of the atom in electromagnetic field of CNT, the adsorbed polarized state of the atom at the surface of CNT $\Psi_p^{ad}(\vec R;t)$, the adsorbed ion state $\Psi_{ion}^{ad}(\vec R;t)$ (ions created by electron tunneling), the desorbed ion state $\Psi_{ion}^{des}(\vec r;t)$, and the desorbed neutral state $\Psi^{des}(\vec r;t)$. Further we represent Hamiltonian (\ref{1.4.14}) in a second quantization representation with respect to the creation and annihilation operators of corresponding states of atoms, while taking into account the structure of a wave function. In this case the basic parameters of description are nonequilibrium average values of density $\langle\hat {n}_\alpha(\vec r)\rangle^t=n_\alpha(\vec r;t),\, \hat {n}_\alpha(\vec r)=\hat {b}^+_\alpha(\vec r)\hat{b}_\alpha(\vec r)$ determines the density operator of atoms in a state $\alpha$. Therefore for macroscopic description of nonequilibrium processes of diffusion, polarization, adsorption, ionization, and desorption in a ``gas-CNT'' system, the set of average densities are chosen for the basic parameters: the average density of the number of atoms $n_a(\vec r;t)$, the average density of the number of polarized atoms in electromagnetic field of CNT $n_a^p(\vec r;t)$, the average density of number of adsorbed polarized atoms $n_a^p(\vec R;t)$, the average density of the number of adsorbed ions $n^+_a(\vec R;t)$ due to electron tunneling, the average density of the number of ions $n^+_a(\vec r;t)$ desorbed from the surface of CNT due to electron tunneling. These parameters of description satisfy a set of transport equations that we obtain by means of nonequilibrium statistical theory of reaction-diffusion processes \cite{rd1, rd2, rd3}. In states of constant transport kernels a set of transport equations has the structure

\bea\label{transE}
\frac{\partial}{\partial t}n_a(\vec r,t)&=&-\nabla_r\cdot(n_a(\vec r,t)\vec v_a(\vec r,t))-K_p n_a(\vec r,t)n^p_a(\vec r,t),\\\nonumber
\frac{\partial}{\partial t}n^p_a(\vec r,t)&=&-\nabla_r\cdot(n^p_a(\vec r,t)\vec v^p_a(\vec r,t))-K_p n_a(\vec r,t)n^p_a(\vec r,t)\\\nonumber
&-&K_{ad} n^p_a(\vec r,t)n^p_a(\vec R,t),\\\nonumber
\frac{\partial}{\partial t}n^p_a(\vec R,t)&=&-\nabla_R\cdot(D^p_{aa}(\vec R,t)\nabla_R)\cdot n^p_a(\vec R,t)-K_{ion} n^+_a(\vec R,t)n^p_a(\vec R,t)\\\nonumber
&+&K_{ad} n^p_a(\vec r,t)n^p_a(\vec R,t),\\\nonumber
\frac{\partial}{\partial t}n^+_a(\vec R,t)&=&-\nabla_R\cdot(D^p_{aa}(\vec r,t)\nabla_R)\cdot n^+_a(\vec R,t)+K_{ion} n^+_a(\vec R,t)n^p_a(\vec R,t)\\\nonumber
&-&K_{des} n^+_a(\vec r,t)n^+_a(\vec R,t)-Z_+e\nabla_R\cdot D_{aa}^+(\vec R,t)\cdot \vec E_{carb}(\vec R,t),\\\nonumber
\frac{\partial}{\partial t}n^+_a(\vec r,t)&=&-\nabla_r\cdot(n^+_a(\vec r,t)\vec v^+_a(\vec r,t))-Z_+e\nabla_r\cdot D^+_{aa}(\vec r,t)\cdot \vec E_{carb}(\vec r,t)\\\nonumber &+&K_{des} n^+_a(\vec r,t)n^+_a(\vec R,t),
\eea

where $\vec v_a(\vec r,t)$ stands for the average velocity of neutral gas atoms, $K_p$ is the polarization constant for atoms in electromagnetic field of CNT that can be estimated by \cite{66},  $\vec v^p_a(\vec r,t)$ denotes the average velocity of polarized gas atoms, $K_{ad}$ is the adsorption constant for polarized atoms, $D^p_{aa}(\vec R,t)$ denotes the coefficient of inhomogeneity of surface diffusion of adsorbed polarized atoms, $K_{ion}$ stands for the ionization constant for adsorbed polarized atoms due to tunneling processes, $D^+_{aa}(\vec R,t)$ denotes the coefficient of inhomogeneity of surface diffusion of adsorbed ions, $\vec E_{carb}(\vec R,t)$ determines electric field at the CNT surface, $K_{des}$ denotes the desorption constant of ions from the CNT surface, $D^+_{aa}(\vec r,t)$ stands for the coefficient of inhomogeneity of diffusion of desorbed ions in the electric field $\vec E_{carb}(\vec R,t)$. The obtained equations are nonlinear and the set of equations can be solved numerically with the determined constant transport kernels and initial and boundary conditions.
Estimation of the transport kernels $K_p$, $K_{ad}$, $K_{des}$, $K_{ion}$ requires the determined values of activation energy of corresponding processes, which could be obtained within \textit{ab initio} approach that we present in the next section.

\section{A computational study of adsorption of \textit{He, NO} on the surface of SWCNT}

Simulations of adsorption of gas atoms and molecules on the surface of SWCNT are conducted with density functional theory within generalized gradient approximation, plane wave basis set and norm-conserving pseudopotentials as it is implemented in ABINIT program package \cite{ABINIT1, ABINIT2, ABINIT3}.
In order to study the effects of chirality and diameter of SWCNT on their adsorption properties several SWCNT of different chiralities (8,0), (5,5), (8,8) are considered, which are known to be semiconducting and metallic types. A supercell that is used in calculation contains a fragment of SWCNT (of 64, 60, 96 carbon atoms respectively) with its axis aligned along one of the coordinate directions surrounded by a vacuum in two other directions. Due to periodical conditions applied a modeled system presents SWCNT of an infinite length separated by a vacuum. A study of the amount of vacuum that would prevent any interactions between periodical images of SWCNT show that the optimal distances between CNT axis are 13.2\AA, 13.0\AA, 14.5\AA \, for CNT(8,0), CNT(5,5) and CNT(8,8) respectively, while their diameters are calculated to be 6.30\AA, 6.82\AA, 10.88\AA  \, respectively.
Geometry of the modeled structures is relaxed with the residual forces on atoms of less than 1.0e-5 Hartree/\AA. The calculations are conducted with cut-off energy of 30 Hartree.

Several possible localizations of adsorbed He atoms on the surface of SWCNT are considered: on top of a carbon atom, on a bond between two carbon atoms and above a center of a hexagonal cell formed by carbon atoms on the surface. The smallest adsorption energy and corresponding localization states are given in a table at the end of this article (see Table I). For adsorption of NO molecule, relaxation of geometry of the system is conducted for several tolerance values of residual forces. A performed \textit{ab initio} molecular dynamics study shows that a molecule drifts along the surface of CNT with a nitrogen in close proximity (2.9-3.2\AA) to carbon atoms with forces between them bigger than 1.0e-4 Hartree/\AA (see Figure 1).  As NO at electron tunneling distances from CNT acts as a charge acceptor, this considerably affects the width of a band gap for semiconducting SWCNT(8,0). Despite the known underestimation of these quantities by DFT methods with non-hybrid pseudopotentials for the small gap systems, the results of 0.65eV for band gap of pristine SWCNT(8,0) and 0.02eV for a SWCNT(8,0) with adsorbed NO qualitatively prove the significance of the effect. Consequently, CNT resistance is altered, which makes semiconducting SWCNT an excellent material for electrochemical sensing of NO molecules.

\begin{SCfigure}
  \centering
  \caption{Electron density profile of NO adsorbed on SWCNT(8,0).  A slab is made across carbon atoms bonds through a center of a hexagonal cell and an oxygen atom. NO is located with an oxygen atom above a center of a hexagonal cell and a nitrogen atom above a carbon atom at the background.}
   \includegraphics[scale=0.5, keepaspectratio=true]{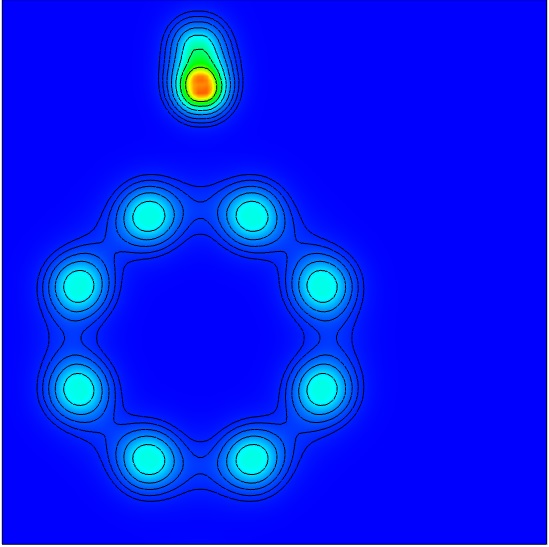}
\end{SCfigure}

Energy of adsorption is calculated as a difference between total energies of the systems according to a formula
\begin{equation}
E_{ads} = E_{SWCNT+ads.gas} - E_{SWCNT} - E_{ads.gas}.
\end{equation}

The dependency of adsorption energy on the distance of gas atoms from CNT surface is shown with respect to CNT chirality (see Figure 2). On the basis of the conducted calculations one can deduce that adsorption energy and adsorption distances of He atom do not depend significantly on the diameter nor chirality of SWCNT.

\begin{figure}[!ht]
\centering
\includegraphics[scale=0.4, keepaspectratio=true]{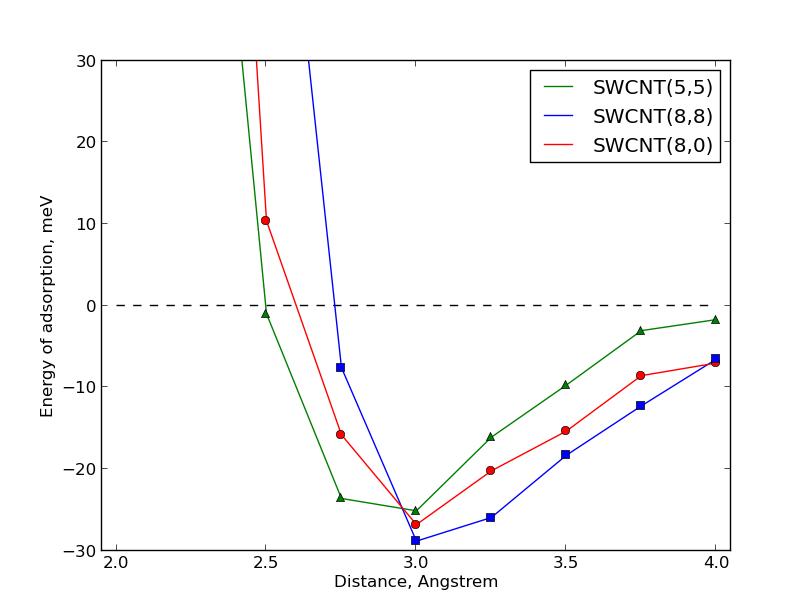}
\caption{Energy of He adsorption vs adsorption distance}
\end{figure}

Adsorption of He on pristine CNT occurs with a very small adsorption energy. This may challenge the developing of a Helium sensor on the basis of pristine SWCNT. An optional solution could present doping of SWCNT with atoms of metals that might increase SWCNT response \cite{ads1,ads2,Pt, abinit, abinit2}. The effect of doping of CNT on adsorption of He atoms is still to be investigated in further research.

In order to study the effect of vacancies in SWCNT structure on gas adsorption the corresponding calculations are carried out. First, it is noted that after removal of one of the carbon atoms from SWCNT the surface in the vicinity undergoes reconstruction, forming a pentagon and one unsaturated binding of carbon atom (see Figure 3) \cite{Maciejewska}.
\begin{figure}[!hb]
\centering
\includegraphics[scale=0.5, keepaspectratio=true]{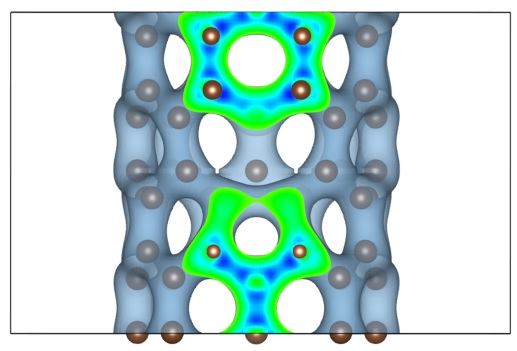}
\caption{Reconstruction of structure as vacancy formation. A slab of electron density}
\end{figure}

This formation has influence on the value of band gap and adsorption properties of material. If an atom of He gas is placed in the vicinity of vacancy the energetically preferable localization of atom is changed, while energy of adsorption remains about the same in comparison to adsorption parameters on pristine CNT. In situations when adsorption of NO molecule occurs on the reconstructed site of CNT, a dangling bond of a carbon atom and unoccupied electron of a nitrogen atom form a chemical bond that leads to functionalization of SWCNT with NO (see Figure 4). Naturally, the value of chemisorption binding energy between NO and SWCNT is much bigger in comparison with adsorption energy (see Table I).

Although chemisorption is significantly easier to detect, a created covalent bond impairs further detection of gases. This demands the techniques of enhancement sensitivity of CNT as ultraviolet illumination \cite{UV}. SWCNT with a rather undestroyed structure with no dangling bindings should be used for gas sensing in order to maintain permanent response of CNT-based nanosensors.

\begin{figure}[!h]
\centering
\includegraphics[scale=0.5, keepaspectratio=true]{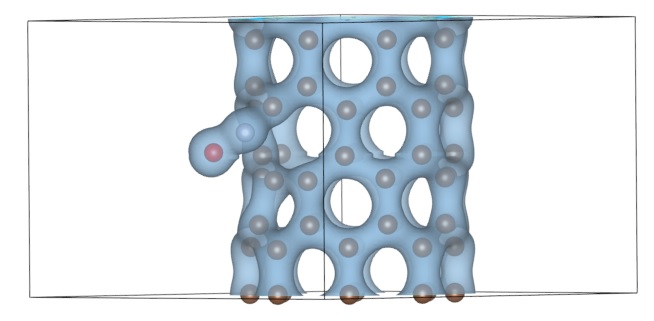}
\caption{Chemisorption of NO on SWCNT(8,0). Electron density within simulated supercell}
\end{figure}

\begin{table}[!ht]
\centering
\caption{Energies of adsorption for various localization states}
\begin{tabular}{| l | m{24mm} | m{26mm} | m{24mm} |}
\hline
Energy of adsorption & cnt(5,5) metallic & cnt(8,0) semiconducting & cnt(8,8) metallic \\
\hline
He     & -25meV & -27meV & -29meV  \\
\hline
localization & top & center & top \\
\hline
He (vacancy) &  -25meV   &  -26meV   & -27meV    \\
\hline
localization & center & center & center \\
\hline
NO           &  repulsive     & repulsive & -11meV \\
\hline
NO (vacancy) & -2.442eV  & -1.112eV   & -2.491eV \\
\hline
\end{tabular}
\end{table}
\newpage
\section{Conclusion}
In an analytical approach for description of adsorption properties of a system ``CNT-electro\-magnetic field-gas'' we employed the Zubarev method of nonequilibrium statistical operator. As a result we derived a set of generalized transport equations for the average values of density of the number of gas atoms above CNT, the number of atoms that are polarized by electromagnetic field of CNT, the number of polarized atoms that are adsorbed on CNT, the number of gas ions that are adsorbed on CNT due to electron tunneling, and the average density of the number of ions desorbed from the surface of CNT due to electron tunneling. The derived equations are nonlinear and can be solved numerically, provided that the transport kernels like polarization constant for atoms in electromagnetic field of CNT, which can be estimated by \cite{66}, adsorption constant for polarized atoms, a coefficient of surface diffusion of adsorbed atoms, ionization constant, and desorption constant are determined.
By means of the \textit{ab initio} technique we simulated adsorption processes of He and NO gases on SWCNT of different sizes and chiralities. The investigations yielded adsorption energies and distances as well as preferable localizations of adsorbed gas atoms. The observed change of band gap of zigzag SWCNT under adsorption of NO enables us to conclude that semiconducting type of SWCNT responds in a more pronounced way on the adsorption of NO, and therefore can be a preferable type of SWCNT as a basis for nanosensors. It was found that the presence of vacancies on SWCNT not only prompts a surface reconstruction, but is also a reason for chemisorption of NO gas molecules and functionalization of CNT.

\section*{Acknowledgements}
Financial support from the National Center for Research and Development under research grant ``Nanomaterials and their application to biomedicine'', contract number PBS1/A9/13/2012 is gratefully acknowledged.

The calculations were conducted with support of Interdisciplinary Center for Computational and Mathematical Modelling (ICM) of Warsaw University, computational grant number G55-19.

\end{document}